# An Error-Surface-Based Fractional Motion Estimation Algorithm and Hardware Implementation for VVC

Shushi Chen[1], Leilei Huang[2,*], Jiahao Liu[1], Chao Liu[1], Yibo Fan[1]

[1]School of Microelectronics, Fudan University, Shanghai, China

[2] School of Integrated Circuits, East China Normal University, Shanghai, China

Email: llhuang@cee.ecnu.edu.cn, fanyibo@fudan.edu.cn

*Abstract*—Versatile Video Coding (VVC) introduces more coding tools to improve compression efficiency compared to its predecessor High Efficiency Video Coding (HEVC). For inter-frame coding, Fractional Motion Estimation (FME) still has a high computational effort, which limits the real-time processing capability of the video encoder. In this context, this paper proposes an error-surface-based FME algorithm and the corresponding hardware implementation. The algorithm creates an error surface constructed by the Rate-Distortion (R-D) cost of the integer motion vector (IMV) and its neighbors. This method requires no iteration and interpolation, thus reducing the area and power consumption and increasing the throughput of the hardware. The experimental results show that the corresponding BDBR loss is only 0.47% compared to VTM 16.0 in LD-P configuration. The hardware implementation was synthesized using GF 28nm process. It can support 13 different sizes of CU varying from 128×128 to 8×8. The measured throughput can reach 4K@30fps at 400MHz, with a gate count of 192k and power consumption of 12.64 mW. And the throughput can reach 8K@30fps at 631MHz when only quadtree is searched. To the best of our knowledge, this work is the first hardware architecture for VVC FME with an interpolation-free strategy.

## I. Introduction

Versatile Video Coding (VVC) [1] is a new video coding standard developed by the Joint Video Experts Group (JVET) established by the ISO/IEC Moving Picture Experts Group (MPEG) and the ITU-T Video Coding Experts Group (VCEG). VVC provides a significant improvement in compression capability over previous generations of standards. To achieve this, VVC introduces more coding tools compared to its predecessors. Even though VVC has adopted many new inter-frame tools, fractional motion estimation (FME) is still the most basic and important one.

In FME hardware design, iteration-free algorithms are preferred since their hardware-friendly characteristic. In VVC Test Model (VTM) [2], the FME process still uses the same two-stage search algorithm as the previous coding standard, i.e., 1/2 followed by 1/4 pixel precision search which is a very classical fast algorithm that provides a good compromise between coding complexity and performance in software. But it needs to search up to 17 sub-pixel points and its multiple iterative processes are not conducive to hardware implementation. Decoder side motion vector refinement (DMVR) [3] uses the error surface to perform FME, thus avoiding a lot of computation in VVC. Reference [4] only searches 12 sub-pixel points by utilizing the result of the integer motion estimation (IME) process. Although it proposes a bilinear quarter pixel approximation scheme, multiple parallelisms and the Hadamard calculation still make the hardware cost up to 1183k gates. Reference [5] reduces the number of sub-pixel points based on the statistical observation of whether the MVD is zero or not. The strategy reduces the computational complexity, but it still requires multiple iterations. Reference [6] introduces a hardware-friendly FME algorithm for VVC. It analyzes the possibility of sub-pixel points being selected as the best prediction points and proposes a hardware-friendly algorithm with different FME search patterns. It avoids the iterative process in motion search, but still requires interpolation calculations, and does not discuss how to obtain a motion vector predictor (MVP). Actually, in hardware, MVP is always determined based on the Rate-Distortion (R-D) cost in the final mode decision [7], which causes MVP to be unknown at the FME stage. Therefore, it is also very important to estimate the MVP accurately in the FME hardware design. Reference [8] proposes a two-stage IME. The first stage only uses the prediction unit (PU) in 4×4 size, and a more accurate MVP is obtained in the second stage. But it is designed for the IME process, which is difficult to apply to the FME process.

To solve the problem of iteration and MVP estimation, this paper proposes an FME algorithm based on the error surface. The major contributions of this paper are as follows:

- An interpolation-free FME approach is proposed based on the error-surface algorithms, which does not need multiple iterations.
- Coarse motion vector prediction (CMVP) is proposed to solve the dependence of estimating MVP in the FME stage.
- An efficient and low-cost FME hardware architecture is designed, to the best of our knowledge, which is the first FME hardware work with an interpolation-free strategy for the VVC standard.

The paper is organized as follows. Section II presents a search algorithm based on error surface and CMVP. Section III introduces the hardware architecture of this algorithm. Section IV shows the experimental result and discussion. Section V concludes the paper.

## II. Hardware-Friendly FME Search Algorithm Based on Error Surface

### A. Error surface on R-D cost

The FME process of VTM is a two-step search algorithm, as shown in Fig. 1(a). However, the dependency between the first and second steps is not conducive to hardware implementation. Therefore, this paper performs the FME process by finding the lowest point of the error surface established around the integer motion vector (IMV) generated by the IME process, which has no data dependence, as shown in Fig. 1(b).

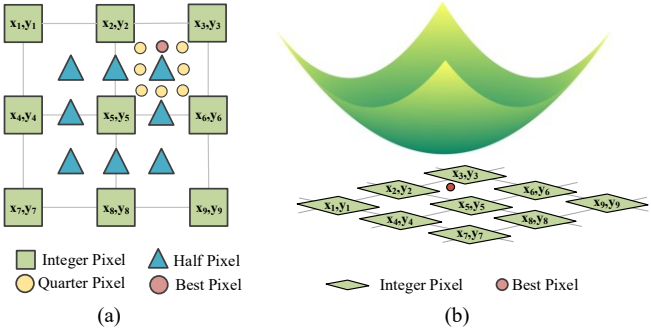

Fig. 1. (a) Two-step FME in VTM. (b) The error surface.

The error surface for FME can be constructed using a five-parameter [9], six-parameter [10], or nine-parameter model [11]. Different parametric models have no significant effect on coding performance [9]. To achieve a compromise between coding performance and complexity, this paper chooses the six-parameter model. Equation (1) is the error surface equation, where the six parameters are represented by $P_1 \sim P_6$, and $C(x_i, y_i)$ denotes the R-D Cost at the point $(x_i, y_i)$, with the best IMV as the coordinate origin.

$$C(x_i, y_i) = P_1 x_i^2 + P_2 y_i^2 + P_3 x_i y_i + P_4 x_i + P_5 y_i + P_6 \quad (1)$$

Solving for the six parameters in (1) requires the cost and location information of at least six points. To cover a wider range and obtain a more accurate solution, the IMV and its eight neighbors are plugged into the error equation to establish the linear equations shown in (2). The overdetermined equations need to be solved by the least square method. And considering that the parameter $P_6$ is a constant that does not affect the location of the extreme value points, only five parameters need to be calculated.

$$\begin{bmatrix} C_1 \\ C_2 \\ C_3 \\ C_4 \\ C_5 \\ C_6 \\ C_7 \\ C_8 \\ C_9 \end{bmatrix} = \begin{bmatrix} x_1^2 & y_1^2 & x_1 y_1 & x_1 & y_1 & 1 \\ x_2^2 & y_2^2 & x_2 y_2 & x_2 & y_2 & 1 \\ x_3^2 & y_3^2 & x_3 y_3 & x_3 & y_3 & 1 \\ x_4^2 & y_4^2 & x_4 y_4 & x_4 & y_4 & 1 \\ x_5^2 & y_5^2 & x_5 y_5 & x_5 & y_5 & 1 \\ x_6^2 & y_6^2 & x_6 y_6 & x_6 & y_6 & 1 \\ x_7^2 & y_7^2 & x_7 y_7 & x_7 & y_7 & 1 \\ x_8^2 & y_8^2 & x_8 y_8 & x_8 & y_8 & 1 \\ x_9^2 & y_9^2 & x_9 y_9 & x_9 & y_9 & 1 \end{bmatrix} \begin{bmatrix} P_1 \\ P_2 \\ P_3 \\ P_4 \\ P_5 \\ P_6 \end{bmatrix} \quad (2)$$

For convenience, the result can be expressed in the matrix form shown in (3).

$$\boldsymbol{P} = (\boldsymbol{X}^T \boldsymbol{X})^{-1} \boldsymbol{X}^T \boldsymbol{C} \quad (3)$$

where $\boldsymbol{P}$ represents five parameters, $\boldsymbol{X}$ is the coordinate of nine points and $\boldsymbol{C}$ is the corresponding cost. Using the five solved parameters to construct the error surface, the extreme value point can be found to obtain the FMV, as shown in (4).

$$\begin{cases} FMV_x = (2P_2 P_4 - P_3 P_5)/(P_3^2 - 4P_1 P_2) \\ FMV_y = (2P_1 P_5 - P_3 P_4)/(P_3^2 - 4P_1 P_2) \end{cases} \quad (4)$$

By solving the error surface equation, the best FMV can be obtained without interpolation.

*B. Estimation of R-D cost*

Rate distortion optimization [12] is widely employed to find optimal coding parameters, as shown in (5).

$$J = D + \lambda * R \quad (5)$$

where $J$ is the R-D cost and $\lambda$ is the Lagrange multiplier. Among the motion estimation process in video coding, the $D$ is usually the sum of absolute difference (SAD) or the sum of absolute transformed difference (SATD) of residuals. The $R$ is usually represented by the bits of the motion vector difference (MVD).

For distortion calculation, SAD is simple but it only describes the difference of the residuals in the temporal domain. On the other hand, SATD is calculated from the Hadamard transformed residuals, which implements a simple time-frequency transformation. Compared to SAD, SATD has higher computational complexity but is more beneficial to improve the Bjøntegaard delta bit rate (BDBR). Therefore, this work uses SATD as a distortion metric.

MVD is the difference between MV and MVP, and MVP is obtained by advanced motion vector prediction (AMVP) in video coding, which requires the exact partition. However, in the hardware, the partition result is unknown in the FME stage. Therefore, this paper proposes CMVP to generate an approximate MVP without considering the exact partition. Taking Fig. 2 as an example, the black lines indicate the actual partition. For the yellow CU, AMVP derives its MVP from the green blocks, while CMVP uses blue blocks' MV which belongs to 8×8 CUs, so MVD can be calculated in the FME stage with CMVP. Moreover, this paper proposes an interlaced processing schedule to implement CMVP hardware, which is described in Section III. D.

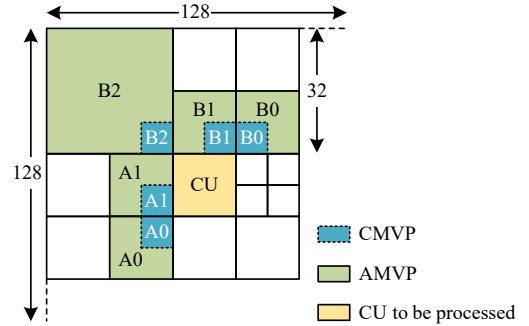

Fig. 2. CMVP and AMVP process in a CTU.

### III. HARDWARE ARCHITECTURE

*A. Overview of the Hardware Architecture*

The overall FME architecture is shown in Fig. 3. It consists of three main parts, which are the cost calculator, the FMV calculator, and the control part. Original and prediction pixels are fed into the cost calculator which generates nine R-D costs of IMV and its eight neighbors. The FMV calculator can solve the parameters of the error surface according to the nine R-D costs and calculate the final FMV. And the control part can generate control signals to control the operation state of the whole system.

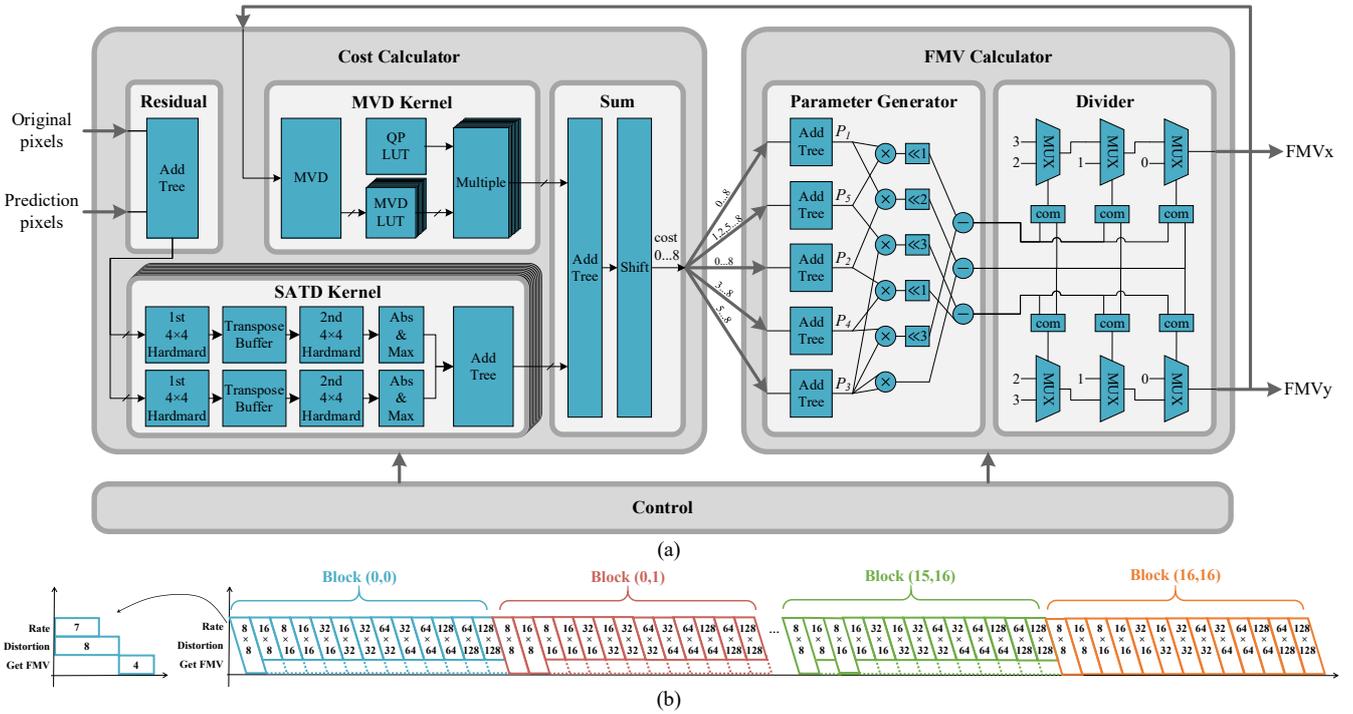

Fig. 3. (a) Hardware architecture of proposed algorithm. (b) Interlaced Processing Schedule.

Moreover, to balance hardware throughput and search efficiency, the minimum CU size supported by this FME architecture is 8×8, and larger CUs are divided into several 8×8 blocks for processing.

*B. Detailed Implementation of the cost calculator*

In the cost calculator, where the final R-D cost is generated by the Sum module, SATD kernels are used to estimate distortion and the MVD kernel is used to estimate the rate.

To improve throughput, the architecture uses nine SATD kernels running in parallel. Each SATD kernel has a throughput of 8×1 pixels per cycle, so it takes 8 cycles to finish an 8×8 block. The 4×4 Hardmard submodules are used to calculate and accumulate the distortion of each 8×8 block. And register-based transpose buffers are used between two steps of Hadamard transformation, so the transposition process is carried out in a "duplex" way.

The MVD kernel can store the best MVs of 8×8 CUs which are adjacent to the current CU. According to the CMVP shown in Fig. 2, the MVP of the current CU can be generated with the stored MVs. The rate of MVD is calculated by the lookup table based on the MVP.

*C. Divider-free FMV Calculator*

In the FMV calculator, the parameter generator can generate the parameters of the error surface according to nine shifted R-D costs. Then the FMV can be calculated in the divider part according to (4). However, there are two division operations, and the fixed-point divider will consume significant hardware resources. Fortunately, the MV is 1/4 pixel precision in FME of VVC, which makes it possible to eliminate the divider. Therefore, the FMV can be rounded to 1/4 pixel accuracy by comparing the 8× numerator to the 3× or 5× denominator. In this way, the use of dividers can be avoided.

*D. Interlaced Processing Schedule*

In this architecture, the 8×8 blocks of different CUs are processed in the interlaced zig-zag scan order, which is convenient to support the CMVP mentioned in Section III. B. The processing schedule is shown in Fig. 3(b). The "(0,0)", "(0,1)", "(15,16)" and "(16,16)" mean the coordinate of different 8×8 blocks. These blocks proceed in zig-zag scan order. In a 128×128 coding tree unit (CTU), there are 256 8×8 blocks in total. The "8×8", "16×8" and "8×16" represent the CUs to which the 8×8 blocks belong. For different CUs, their 8x8 blocks are scrambled and carried out in an interlaced order. For the first 8×8 block which belongs to 8×8 CU, it takes 12 cycles to output FMV. Then the latter 8×8 blocks can be processed in the pipeline. As for CUs larger than 8×8, only the last 8×8 block in CU needs to calculate FMV, while other 8×8 blocks before it only calculate the rate and distortion, as shown by the dotted line in Fig. 3(b).

IV. EXPERIMENTAL RESULTS AND COMPARISON

*A. Software Performance*

The proposed algorithm is implemented in VTM 16.0 and tested under the Common Test Conditions (CTC) [13]. Intel Core i7-860 CPU (3.8 GHz), 16 GB RAM, and Ubuntu 16.04 LTS are used to perform the test. The test sequences include Class B (1080p), C (WVGA), D (QWVGA), E (720p), and F (Screen Content), and each sequence was tested for 100 frames.

Table I gives the encoding efficiency of each sequence under LD-P configuration. As shown in the table, the average BDBR of the proposed algorithm increases by only 0.472%. The previous study on error surfaces [9] did not consider the effect of distortion calculation on prediction accuracy. It uses SAD for distortion calculations and performs a further search around the obtained FMV. Reference [6] performs FME with a defined pattern and its BDBR increases by 0.344%.

TABLE I. CODING EFFICIENCY COMPARISON

| | Sequence | VCIP'16 [9] HEVC,software BDBR | MMSP'21 [6] VVC,software BDBR | VLSI'15 [4] HEVC,hardware BDBR | This Work VVC,hardware | | | |
|---|---|---|---|---|---|---|---|---|
| | | | | | BDBR(Y) | BDBR(U) | BDBR(V) | BDBR |
| Class B | MarketPlace | | \ | \ | 0.167 | 0.855 | -0.036 | 0.207 |
| | BasketballDrive | | 0.45 | 2.44 | 0.321 | 0.269 | 0.538 | 0.342 |
| | BQTerrace | 0.28 | 0.21 | 2.10 | 0.434 | 1.702 | 1.979 | 0.629 |
| | Cactus | | 0.16 | 1.78 | 0.070 | -0.027 | 0.429 | 0.105 |
| | RitualDance | | \ | \ | 0.187 | 0.212 | 0.538 | 0.222 |
| Class C | BasketballDrill | | 0.30 | \ | 0.676 | 0.068 | 0.942 | 0.649 |
| | BQMall | 0.32 | 0.36 | \ | 0.609 | 0.739 | 0.398 | 0.598 |
| | PartyScene | | 0.38 | \ | 0.815 | 1.036 | 0.501 | 0.807 |
| | RaceHorsesC | | 0.54 | \ | 0.600 | 0.336 | 1.179 | 0.627 |
| Class D | BasketballPass | | 0.69 | \ | 0.554 | 0.472 | 1.368 | 0.624 |
| | BlowingBubbles | 0.57 | 0.45 | \ | 0.441 | 0.436 | 0.587 | 0.454 |
| | BQSquare | | 0.52 | \ | 0.990 | 0.635 | 1.256 | 0.986 |
| | RaceHorses | | 0.77 | \ | 0.763 | 0.330 | 0.157 | 0.669 |
| Class E | FourPeople | | \ | \ | 0.330 | -1.282 | 0.183 | 0.150 |
| | Johnny | 0.17 | \ | \ | 0.599 | 0.545 | 1.122 | 0.622 |
| | KristenAndSara | | 0.31 | \ | 0.225 | -0.028 | 0.856 | 0.264 |
| Class F | ArenaOfValor | | \ | \ | 0.213 | 0.103 | 0.119 | 0.192 |
| | SlideEditing | 0.38 | 0.80 | \ | 0.033 | -0.093 | 0.012 | 0.016 |
| | SlideShow | | \ | \ | 0.358 | 0.545 | 1.679 | 0.482 |
| | BasketballDrillText | | \ | \ | 0.992 | -0.182 | -0.129 | 0.750 |
| Average | | 0.344 | 0.457 | 2.10 | 0.469 | 0.334 | 0.684 | 0.472 |

* VLSI'15 [4] and VCIP'16 [9] have tested more sequences, but for the fairness of the comparison, the table only lists the results for the shared sequences.

Noteworthy, only reference [6] and our work were performed on VVC. Compared with this work, references [9] and [6] still require interpolation, and there is no corresponding hardware implementation. Reference [4] relies on IME results to reduce the number of points to be searched, and only three sizes of CU can be searched. It has corresponding hardware, but the BDBR increases by up to 2.10%.

To sum up, our proposed algorithm has a similar performance to the software solution, and there is a corresponding hardware implementation.

*B. Hardware Performance*

The proposed architecture is synthesized using the Design Compiler to measure the timing, power and area. All synthesis results are obtained under the GF 28 nm cell library.

The proposed hardware can support searching up to 13 CU sizes. As shown in Fig. 3(b), it takes 12 cycles to output the best motion vector of the first 8×8 CU, and other CUs are pipelined. So, the number of cycles to process a CTU is (12-8)+8×13×(128×128)/(8×8) = 26628 cycles. When the hardware design works at 400MHz, it can achieve 4K@30fps throughput. If only quadtree is considered, there are only five CU sizes. In this case, the design has a throughput of 8K@30fps at 631MHz. Actually, the maximum frequency of this hardware can reach 800MHz under GF 28nm process.

Table II shows a comparison of different FME hardware architectures in the literature. Reference [14] can search for all 48 FMVs near IMV, but this also increases computational complexity. It can achieve 4K@120fps at 800MHz, but this is only for 8×8 PU and does not support other PU sizes. The design proposed in [4] uses a search pattern based on a bilinear quarter pixel approximation, which helps reduce area and power consumption but reduces its performance. Although it achieves a throughput rate of 8K@30fps, it can only support three sizes of CU from 64×64 to 16×8/8×16. On the other hand, our implementation can support 13 sizes of CU from 128×128 to 8×8.

V. CONCLUSION

This paper introduces an error-surface-based FME algorithm to avoid the interpolation process and implements the corresponding hardware. To the best of our knowledge, this is the first hardware design for VVC FME with an interpolation-free strategy. In the VVC standard, the error surface is also used in DMVR for FME. Therefore, our future work will consider the deep integration of DMVR and FME to support more tools at the lowest possible cost.

ACKNOWLEDGMENT

This work was supported in part by the National Natural Science Foundation of China under Grant 62031009, in part by Alibaba Innovative Research (AIR) Program, in part by the Fudan-ZTE Joint Lab, in part by CCF-Alibaba Innovative Research Fund For Young Scholars.

TABLE II. HARDWARE COMPARISON

| Design | VLSI'15 [4] | ISCAS'18 [14] | This Work | |
|---|---|---|---|---|
| Standard | HEVC | HEVC | VVC | |
| Technology | 65nm | 45nm | 28nm | |
| CU size | 64×64~8×16 | 8×8 | 128×128~8×8 | |
| Interpolation | necessary | necessary | unnecessary | |
| Gate | 1183k | / | 192k | |
| Area | / | 100733.4μm² | 89925.26μm² | |
| Max Freq. | 188MHz | 800MHz | 800MHz | |
| Throughput | 8K@30fps | 4K@120fps | 4K@30fps @400MHz | 8K@30fps* @631MHz |
| Power | 198.6mW | 45.07mW | 12.64mW | 26.45mW |
| CU Number | 3 | 1 | 13 | 5 |
| BDBR | 2.10% | / | 0.469% | 0.516% |

*For the result of 8K@30fps, anchor and our work only searched 5 sizes of CU under quadtree partitioning.